\documentstyle[12pt,aaspp4]{article}
\tighten

\begin{document}

\title {Mid-infrared Imaging of a Circumstellar Disk Around HR 4796:\\
Mapping the Debris of Planetary Formation}

\author {D.W. Koerner\altaffilmark{1, }\altaffilmark{2}, M.E. 
Ressler\altaffilmark{1}, M.W. Werner\altaffilmark{1}, and D.E. 
Backman\altaffilmark{3}}
\altaffiltext{1} 
{Jet Propulsion Laboratory, California Institute of Technology,
4800 Oak Grove Dr., Pasadena, CA 91109}
\altaffiltext{2}
{University of Pennsylvania, 4N14 DRL, 
209 S. 33rd St., Philadelphia, PA 19104-6396} 
%{\it davidk@laplace.physics.upenn.edu}}
\altaffiltext{3}
{Franklin \& Marshall College, Dept.\ of Physics \& Astronomy, 
PO Box 3003,
Lancaster, PA 17604-3003} %

\bigskip
\bigskip

\begin{abstract}

We report the discovery of a
circumstellar disk around the young A0 star, 
HR 4796, in thermal infrared imaging carried out at the W.M. Keck Observatory.
By fitting a model of the emission from a flat dusty disk
to an image at $\lambda$~=~20.8~$\mu$m, we derive a disk inclination, 
$\iota$~=~${72^\circ}^{+6^\circ}_{-9^\circ}$ from face on,
with the long axis of emission at PA ${28^\circ}\pm6^\circ$.
The intensity of emission does not decrease with radius as expected 
for circumstellar disks but {\it increases} outward from the star, 
peaking near both ends of the elongated structure.  We simulate this
appearance by varying the inner radius in our model and find 
an inner hole in the disk with radius R$_{in}$ = 55$\pm$15 AU. 
This value corresponds to the radial distance of 
our own Kuiper belt and may suggest a source of dust
in the collision of cometesimals. By contrast with the appearance
at 20.8 $\mu$m, excess emission at $\lambda$ = 12.5 $\mu$m is faint and
concentrated at the stellar position. Similar emission is also detected 
at 20.8 $\mu$m in residual subtraction of the best-fit model from the image.
The intensity and ratio of flux densities at the two wavelengths 
could be accounted for by a tenuous dust component that is confined 
within a few AU of the star with mean temperature 
of a few hundred degrees K, similar to that
of zodiacal dust in our own solar system.
The morphology of dust emission from 
HR 4796 (age 10 Myr) suggests that its disk
is in a transitional planet-forming 
stage, between that of massive gaseous proto-stellar disks  
and more tenuous debris disks such as the one detected 
around Vega.

\end{abstract}

{}

\section {Introduction}

It is now well over a decade since coronagraphic imaging of stellar light 
scattered by a disk around $\beta$ Pictoris provided 
circumstantial evidence for the presence of an extra-solar planetary 
system (Smith \& Terrile 1984). Since then, coronagraphic and adaptive-optics 
(AO) searches have continued the quest for  
additional examples (Smith, Fountain, \& Terrile 1992; Kalas \& Jewitt 1995), 
but none have been successful at imaging light scattered by dust around
similar IRAS-selected candidates (see Table VII from Backman \& Paresce 1993, 
for example). This failure was originally interpreted to suggest that the 
disk around $\beta$ Pic is highly unusual (Smith et al.\ 1992),
but subsequent identification of a few other main sequence A stars with  
fractional infrared excesses of the same magnitude 
has been used to argue that $\sim$20\%
of all A-type stars pass through an early phase like that of $\beta$ Pic
(Jura et al.\ 1998). If true, the failure
to image other candidates is likely due to their greater distance;
limitations in current coronagraphic techniques would prevent detection
of even $\beta$ Pic's disk if it were several times farther away 
(cf. Kalas \& Jewitt 1996). An alternate approach is suggested by 
successful imaging of the thermal dust emission from the 
disk around $\beta$ Pic at mid-infrared
wavelengths (Lagage \& Pantin 1994; Pantin, Lagage, \& Artymowicz 1997). 
As part of an effort to take full advantage of this technique, 
we present thermal infrared imaging of the circumstellar environment of 
HR 4796, a more distant A0 star (Houk 1982) thought to have a disk
similar to that of $\beta$ Pic on the basis of its IRAS excess (Jura 1991;
Jura et al.\ 1993; 1995; 1998).

At a distance of 67$\pm$3 pc ($\pi$ = 14.91$\pm$0.75 mas; Hipparcos
Catalog), HR 4796 is three and a half times farther away 
than $\beta$ Pic (d = 19.3$\pm$0.2 pc; $\pi$ = 51.87$\pm$0.51 mas), 
but exhibits a strong IRAS excess with optical depth  
estimated to be twice as large (Jura 1991).
An M dwarf companion is detected
7.7$''$ ($\sim$ 500 AU) from the star and is believed to be 
physically bound on the basis of common proper motion (Jura et al.\ 1993). 
The age of the system is estimated to be $t$ = 10 Myr on the basis of
isochrone fitting (Stauffer et al.\ 1995;
Jura et al.\ 1998), in keeping with its
identification as an outlying member of the Centaurus-Lupus association 
(Jura et al.\ 1993) with average age $t$ = 12-15 Myr 
(de Geus, De Zeeuw, \& Lub 1989). This age is younger
than other well-known members of the Vega class ($t \sim 100$ Myr)
and suggests that HR 4796 may be somewhat transitional between young
stars ($t \sim 1$ Myr) with optically thick disks 
(Koerner \& Sargent 1995; Mannings, Koerner, \& Sargent 1997) 
and ``Vega-type'' stars (cf.\ Backman \& Paresce 1993).

\section{Observations and Results}

We observed HR 4796 with JPL's mid-infrared camera MIRLIN at the F/40
bent-Cassegrain focus of the Keck II telescope on UT 16 March 1998.
MIRLIN employs a Boeing 128$\times$128 pixel, high-flux Si:As BIB detector
with a plate scale at Keck II of 0.14$''$ per pixel. The latter was 
determined  by scanning a star across both MIRLIN's field of view and that of
the Keck II guide camera for which the plate scale has been 
measured. The associated field of view was 17.5$''$. Background subtraction
was carried out by chopping the secondary mirror at a 4 Hz rate with
8$''$ throw in the north-south direction and by nodding the telescope a
similar distance in the east-west direction after coadding a few hundred chop
pairs. Observations were
carried out in filters centered at $\lambda$ = 12.5, 20.8, and 24.5 $\mu$m 
with widths 1.2, 1.7, and 0.8 $\mu$m and for on-source integration 
times 6, 17, and 6 minutes respectively.  
Small dither steps were taken between chop-nod cycles. 
Source images on the double-differenced frames were  
shifted and added to make the final 64 $\times$ 64 
(8.3$''$ $\times$ 8.3$''$) images. 
Infrared standards $\alpha$ Sco, $\sigma$ Sco, and $\alpha$ Lyr were
observed in the same way at similar airmasses. An additional sequence of
images was obtained by alternating rapidly between the 12.5 and 20.8 $\mu$m
filters with no telescope offsets in order to accurately register images
at both wavelengths.

A two-color rendition of the emission from HR 4796 at 
$\lambda$ = 12.5 and 20.8 $\mu$m is displayed in Fig.\ 1. 
The underlying images are shown separately in Fig.\ 2a and 2b.
Peak 12.5 $\mu$m emission is compact and
centered on an elongated source at 20.8 $\mu$m.
At $\lambda$ = 12.5 $\mu$m, peak flux density of 
184$\pm$18 mJy comprises nearly all of the point-like emission rising above a
low-level plateau.  Fainter emission accounts for an additional 40 mJy and
is slightly extended to the NE and SW. We estimate the stellar photospheric
emission to be 122 mJy at this wavelength by comparison with flux densities
for $\alpha$ Lyr at $lambda$ = 10 $\mu$m (Tokunaga 1998) and 
assuming similar temperatures, $M_V$ = 5.80 for HR 4796, 
and 0.03 for $\alpha$ Lyr. 
This implies a total excess that is nearly equal to the photospheric 
emission; half of this arises from very near 
the star. In the corresponding image at 20.8 $\mu$m, emission is elongated 
$\sim$3$''$ at PA 30$^\circ$ with total flux
density 1.88$\pm$0.2 Jy, consistent with the 20 $\mu$m bolometer measurement of
Jura et al.\ (1993) (1.86 Jy). Measurements at $\lambda$ = 24.5 $\mu$m
look roughly similar and have total flux density 2.27$\pm$0.7 Jy.
Surprisingly, the center of the 20.8 $\mu$m structure is not coincident 
with the peak emission but lies between bright peaks located at each end.
Below, we interpret this morphology as arising from 
a nearly edge-on circumstellar disk with an inner hole
and investigate its properties with the aid of a 
simple model.

\section {Modeling and Discussion}

Our measured flux densities are listed in Table I and plotted in Figure 3
together with color-corrected IRAS measurements and an upper limit obtained 
with the JCMT at $\lambda$ = 800 $\mu$m by Jura et al. (1993). The
spectral distribution of the flux densities exhibits a nearly 
black-body shape which peaks at $\lambda$ $\approx$ 60 $\mu$m, well
away from the maximum stellar photosphere emission. It can be reproduced by
emission from a model disk with mass approximately 1.0 M$_\oplus$
which is largely devoid of radiating particles inside a radius of a 
few times 10 AU (Jura et al.\ 1995; 1998). In order to uniquely constrain
the spatial distribution of dust, however, it is desirable to gain the 
maximum amount of information from images  
which resolve the circumstellar material.

The optically thin radiation from an annulus of width $dr$ and radius $r$
in a flat dusty circumstellar disk is estimated by  
Backman, Gillett, \& Witteborn (1992) to be
$$ f(r) = \tau_{r_0} \biggl( { r \over r_0 } \biggr )^\gamma 
\varepsilon_\lambda \  B[T_p(r),\lambda] \biggl( {2 \pi r dr \over D^2} sr 
\biggr ) Jy, $$

\noindent
where $\tau_{r_0}$ is the geometrical optical depth perpendicular to the 
disk plane at a fiducial radius $r_0$, $T_p$ is the particle temperature, 
$\varepsilon_\lambda$ is
the particle radiative efficiency relative to a black body, 
and $D$ the 67 pc distance to HR 4796. For a distribution of grains 
with effective size, $\lambda_0$,
we assume $\varepsilon_\lambda = 1$ for $\lambda < \lambda_0$ and 
$\varepsilon_\lambda = (\lambda/\lambda_0)^{-1}$ for  $\lambda > \lambda_0$
(see Appendix D of Backman
et al.\ (1992) for the relation between  $\lambda_0$ and a
particle size distribution).
The grain temperature is then 
$T_p(r) = 468 (L_*/\lambda_0)^{0.2}(r/1{\rm AU})^{-0.4}$ where $L_*$ is the 
stellar luminosity in solar units.  
%For  40--80$\mu$m; we assume $\lambda_0$ = 60 $\mu$m. 
HR 4796 has a temperature similar to $\alpha$ Lyr with 
$M_V$ = 5.80 at D = 67 pc. Then, assuming
L$_*$ = 54 $L_{\odot}$, $M_V$ = 0.03, and D = 8.1 pc for 
$\alpha$ Lyr, we find $L_*$ = 18.1 $L_\odot$ for HR 4796. 
Parameters which remain to be determined include $\lambda_0$,
power-law index of the optical depth $\gamma$, inner and 
outer radii  $r_{in}$ and $r_{out}$, $\tau_{r_0}$,  inclination
angle $\iota$ (from face on), and position angle $\theta$.

The morphology of the 20.8 $\mu$m emission in Fig.\ 2b provides a 
useful constraint on $r_{in}$, $\iota$, $\theta$, and $\tau_{r_0}$
under the assumption of a single power-law structure to the radial
optical depth. However, it does little to constrain $r_{out}$,
$\gamma$, and $\lambda_0$. The latter can be better determined from
fits to the flux density distribution, especially if values of 
the former can be derived from a fit to the image. In both cases,
we choose to evaluate the uncertainties in the range of acceptable
parameter values by calculating the probability of the models
given the data. We use the above formulation to simulate both the image
and the flux density distribution and calculate the reduced $\chi^2$ 
over the range of parameter values. The associated probability of the data
given the model is taken to be $P = e^{-\chi^2}$. The probability 
of the model given the data is then estimated by summing 
and normalizing the probabilities over the whole range of parameter 
space considered. For the estimate of
$\iota$, individual probabilities are scaled by cos$(\iota)$ to account 
for the {\it a priori}
expectation that the system is oriented edge on (see Lay et al.\ 1997
for a more detailed description of this Bayesian approach).

To simulate the image of HR 4796 at 20.8 $\mu$m, we 
calculated the emission from a thin disk as prescribed above at each 
point of a finely sampled grid ($\Delta$s = 0.015$''$) and added
a flux density corresponding to the stellar photosphere to the central
pixel. The result was binned to the resolution of the observations 
and convolved with a normalized image of a
standard star obtained at similar air mass and
within a short time of the HR 4796 observations. The model image
was subtracted from the data to 
derive reduced $\chi^2$ and $P$ in the usual way. Initial 
disk parameter values were estimated from the distribution of flux 
densities in order to determine orientation parameters
$\iota$ and $\theta$. The most likely values of these were then 
taken in a second model fit while varying $r_{in}$, $\tau_{r_0}$,
and $\gamma$.
Assuming the best values from the latter, a 
repeat fit to $\iota$ and $\theta$ was completed as a check. 
The final results for $\iota$ and $\theta$ agreed with the
initially derived values $\iota$ = ${72^\circ}^{+6^\circ}_{-9^\circ}$ 
and $\theta$ = ${28^\circ}\pm6^\circ$.

To test for the presence and size of any clearing in the disk, 
we considered a range of values 
for $\gamma$ and $\tau_{r_0}$, since these also affect the radial 
intensity of emission. The value of 
$\gamma$ was varied between 0 and -2.5; $\tau_{r_0}$ was considered
over the range 10$^{-5} < \tau_{50 AU} < 1$. The appearance of the
20.8 $\mu$m emission is not sensitive to increases in the 
choice of outer radii larger than the boundary of emission in the image, 
since the temperature rapidly decreases to levels at which emission 
is negligible at that wavelength; we adopted $r_{out}$ = 125 AU.
The resulting estimate of $r_{in}$ is evident in the plot in Fig.\ 4. 
It is clear from this figure that the source of 
emission in Fig.\ 2b cannot extend inward to 
within 20 AU of the star as an extrapolation of 
a single power-law profile to the radial optical depth.
In fact, the inner radius of the disk is estimated to be a good
deal larger,  $r_{in}$ = 55$\pm$15AU. 
%The reliability of this result is bolstered by the fact
%that the  30 AU total uncertainty in the value of $r_{in}$ corresponds to the 
%angular resolution of the observations, 0.45$''$, and by the result that
%the choice of power-law index for the geometrical optical depth has
%little effect on the final estimate. 

The most probable model is displayed together with
the residuals in  Fig.\ 2c and 2d. 
It is apparent that, although the presence of a hole leads to
a good match with the outer disk, some emission close to the star
remains in the residual image. This emission looks similar to the 
12.5 $\mu$m image in appearance and may have a common origin in a
warmer population of dust grains near the star with greatly decreased 
optical depth relative to that of the outer disk.
The unresolved flux density in excess of the stellar photosphere 
is 62 mJy at $\lambda$ =  12.5 $\mu$m;  excess residual flux density at
$\lambda$ = 20.8 $\mu$m is 96 mJy. These values suggest mean 
temperatures of 200-300 K and associated distances from the star of 3-6 
AU, depending on whether the grains are the same size as in the outer disk
or radiate as black bodies.
It holds in either case 
that the majority of grains giving rise to the emission 
lie well interior to the 55-AU radius of the hole and in a radial
zone which corresponds to that of the 
zodiacal dust in our own solar system. 

%\subsection{Modeling the Flux Densities with Imaging Constraints}

The properties of the outer disk are 
best estimated with measurements of the emission at far infrared
and millimeter wavelengths which probe cold dust 
(cf.\ Beckwith et al.\ 1990). Assuming an inner radius derived
from our fit to the high-resolution 20.8 $\mu$m image, we
try to estimate these properties with a model of the spectral
distribution of flux densities. In addition to IRAS fluxes and 
an upper limit at $\lambda$ = 800 $\mu$m, our measurements in narrow
bands at $\lambda$ = 12.5, 20.8, and 24.5 $\mu$m provide new constraints
on the distribution of material. Our image at 12.5 $\mu$m
demonstrates further that much of the infrared excess at that wavelength 
does {\it not} arise from the ``disk'' outside $r_{in}$ = 55 AU.
Consequently, we do not attempt to fit either that measurement or 
the IRAS 12 $\mu$m flux density.

We varied four parameters,  $\tau_{50AU}$, $\gamma$, $r_{out}$,
and $\lambda_0$, over the ranges, 10$^{-5}<\tau_{50AU}<1$, -4.5$<\gamma,0$,
56AU$<r_{out}<$200AU, and 10$\mu$m$<\lambda_0<80\mu$m 
in a fit to all flux densities at wavelengths longer than 20 $\mu$m
plotted in Fig.\ 3. Flux from the inner region was derived from a
fit to the ``point-source'' flux densities at 12.5 and 20.8 $\mu$m and
subtracted (89, 88, 23, and 6 mJy at 24.5, 25, 60, and 100 $\mu$m).
%The scatter in the data about the model curve exceeds the formal
%uncertainties quoted for the measurements
%by factors of several. Consequently, it is difficult
%to attach a meaning to the width of the probability distributions for
%each of the parameter values; we do not plot those here.
%If the scatter in the data about the best-fit model is
%taken as a measure of the true uncertainty, 
The preferred parameter values are $r_{out} \approx 55-80$ AU 
and $\lambda_0 \approx 10-40 \mu$m; 
$\tau_{50AU}$ and $\gamma$ are not well determined within these
ranges.  Unless maintained by some orbiting body, an 
extremely narrow ring is probably too short-lived to be likely.
Consequently, we assume $r_{out}$ = 80 AU for the model plotted in
Fig.\ 3, together with 
$r_{in}$ = 55 AU, $r_{out}$ = 80 AU, $\lambda_0$ = 30 $\mu$m, 
$\tau_{50AU}$ = 0.065 and $\gamma$ = -2.0.

\section {Summary and Conclusions}

These observations provide direct evidence for the existence of a disk  
around the young A0 star, HR 4796, with size and orientation that
are corroborated by independent observations taken 2 days later
(Jayawardhana et al.\ 1998). Further, the morphology of emission 
at $\lambda$ = 20.8 $\mu$m reveals an inner hole in the disk with 
radius 55$\pm$15 AU. Emission in
excess above the photosphere at 12.5 $\mu$m arises predominantly
from a region interior to this radius. When compared to excess 20.8 $\mu$m
emission at this location, the 12.5 $\mu$m radiation 
yields a source temperature of a few hundred K, 
corresponding to a radial distance of a
few AU from the central star.  The outer and inner components
of the dust distribution are analogous, respectively, to the Kuiper Belt
and zodiacal components of dust within our own solar system
(Backman et al.\ 1997). Taken together, these properties 
are startlingly like those believed to have existed in 
the late bombardment stage of the early solar system, when collisions 
between cometesimals in the outer solar system and between planetesimals 
or asteroids in a terrestrial planet-forming zone were likely to have
generated a large population of smaller dust grains distributed in much
the same way as those around HR~4796.

\acknowledgments

We wish to thank Robert Goodrich, operator assistants  
Ron Quick and Joel Aycock, and the W.M. Keck Observatory (WMKO) 
summit staff for
helping to adapt and operate MIRLIN at the Keck II visitor
instrument port. Our heart-felt thanks extend especially 
to director F.Chaffee for supporting the use of MIRLIN at WMKO.
We are also indebted to M. Jura for many helpful discussions and to 
R. Jayawardhana and L. Hartmann for sharing results prior to publication.  
Development of MIRLIN was supported by the JPL Director's 
Discretionary Fund and by an SR+T award from NASA's Office 
of Space Science.  The use of MIRLIN at the Keck Observatory 
was supported by an award from NASA's Origins program.  %We 
%thank J.Bock, G.Reyes, and J. Van Cleve (Cornell University) 
%for assisting in the development of MIRLIN. 
Portions of this work were carried
out at the Jet Propulsion Laboratory, California Institute of 
Technology, under contract with the National Aeronautics and
Space Administration.
Data presented herein were obtained at WMKO, 
which is operated as a scientific partnership 
between Caltech, University of California, and NASA, and 
was made possible by the generous financial support of the
W.M. Keck Foundation.

%\vfill
%\eject
% Figure Captions here...

\figcaption{Composite 2-color image of the emission from HR 4796
at $\lambda$ = 12.5 (cyan) and 20.8 $\mu$m (red). Registration was
accomplished by taking images at the same telescope position and
rapidly cycling between both filters.}

\figcaption{a) Image of the intensity of emission from HR 4796 at
$\lambda$ = 12.5. b) As in a, but at 20.8 $\mu$m. c) 
Model of the 20.8 $\mu$m emission from a flat dusty disk with
a 55 AU inner hole and a central star with flux density matched to that
of HR 4796. The model was generated as described in the text and convolved
with a PSF derived from a standard star. d) Residuals from the subtraction 
of the model from the image with exaggerated intensity scale. 
Excess 20.8 $\mu$m emission near the star has a flux density of 96 mJy.}

\figcaption{Plot of the flux densities for HR 4796 listed in Table I.
The solid curve marks the total emission from the stellar photosphere
(dashed line) and a model of the emission from a disk with 55 AU
inner hole (dotted line).}

\figcaption{Plot of the probability distribution for values of the
radius of an inner hole in the disk. As described in the text, the 
radial power-law index of the geometric optical depth was also varied
from $\gamma$ = -2.5 to 0.0. The distribution over all values of
$\gamma$ is shown as a solid line. That for $\gamma$ = 0.0 is plotted 
as a dashed and dotted line, and for $\gamma$ = -2.5 as a dotted line. }

% TABLE2.TEX -- Sample table 2.

\begin{deluxetable}{lccccc}
%\tablewidth{33pc}
\tablewidth{0pc}
\tablecaption{Flux Densities for HR 4796}
\tablehead{
\colhead{$\lambda_{eff}$} & \colhead{$\delta\lambda$} &
\colhead{Flux Density} & \colhead{Uncertainty} &
\colhead{Photosphere} & \colhead{Excess}  \\
\colhead{($\mu$m)} &  \colhead{($\mu$m)} & \colhead{(Jy)} &
\colhead{(Jy)} &  \colhead{(Jy)} & \colhead{(Jy)}  }
\startdata
12.5 &  1.2 & 0.223 & 0.018 & 0.122 & 0.101$\pm$0.018 \nl
20.8 & 1.7 & 1.880 & 0.170 & 0.047 & 1.813$\pm$0.170 \nl
24.5 & 0.8 &  2.270 & 0.700 & 0.033 & 1.994$\pm$0.700 \nl
12.0 & 6.5 &  0.309 & 0.028 & 0.136 & 0.173$\pm$0.028 \nl
25.0 & 11.0 & 3.280 & 0.13 & 0.032 & 3.250$\pm$0.130\nl
60.0 & 40.0 & 8.640 & 0.43 & 0.006 & 8.630$\pm$0.430 \nl
100.0 & 37.0 &  4.300 & 0.34 & 0.002 & 4.300$\pm$ 0.340 \nl
800.0 & 100.0 & $<$ 0.028 & -- & -- & -- \nl
\enddata

\end{deluxetable}

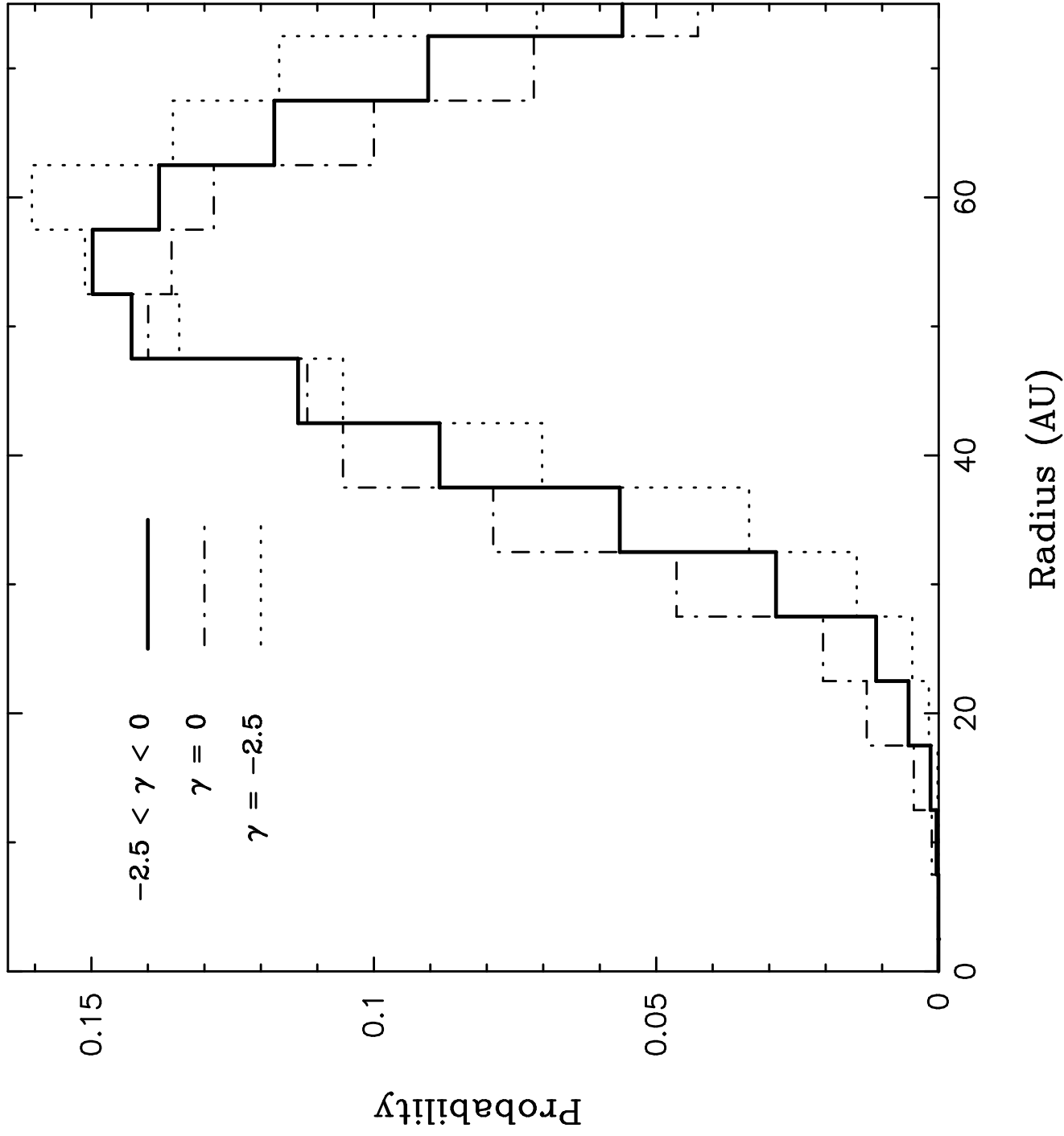
\end{document}